\documentclass[runningheads,a4paper]{llncs}

\usepackage{amssymb}
\usepackage{graphicx}

\usepackage{url}

\newcommand{\keywords}[1]{\par\addvspace\baselineskip
\noindent\keywordname\enspace\ignorespaces#1}

\begin{document}


\title{Integration Checker of JAVA  P2P distributed  System with Auto Source Code Composition }

\titlerunning{Integration Checker by auto source code composition}

%
%
\author{Lican Huang
 \\}

\authorrunning{Lican Huang}

\institute{Zhejiang Sci-Tech University , Domain Search Networking  Technology Co., Ltd,\\
HangZhou, P.R.China, 310018  \\  LicanHuang@zstu.edu.cn; huang\_lican@yahoo.co.uk\\ }

%
%

\toctitle{Lecture Notes in Computer Science}
\tocauthor{Authors' Instructions}
\maketitle

\begin{abstract}

 This paper presents an integrity checker of JAVA  P2P  distributed  system  with auto source code composition. JAVA distributed
 system must guarantee the integrity of program itself and the system components of JAVA virtual machine against attackers, hackers, spies, cheaters, conspirators, etc.  There are lots of trusted computing methods to  guarantee the integrity of the system. We here present a novel method using  just-in-time auto source code composition to generate autocheck class for integrity measure and encrypt  of integrity reporting.
 By   companies'  effort,  we have implemented and use it in DSCloud platform.

\end{abstract}

\section{Introduction}

Distributed systems involve many hardware and software components as well as many users who use the systems. To guarantee the correct results of distributed computing  and  the trusted access of information against attackers is vital important issue. Trusted Computing (TC)  developed and promoted by the Trusted Computing Group\cite{urltcg}  provides strategies to solve the above problem from hardware and software aspects.  With Trusted Computing  enforced by computer hardware and software, the computer  behaves in expected and trustful ways. Systems based on Trusted Computing can protect critical data and systems against a variety of attacks, enable secure authentication and strong protection of unlimited certificates, keys, and passwords that otherwise are accessible, establish strong machine identity and integrity, help satisfy regulatory compliance with hardware-based security, and so on.  The technologies of  Trusted Computing  provide more secure remote access through a combination of machine and user authentication, protect against data leakage by confirmation of platform integrity prior to decryption, provide hardware-based protection for encryption and authentication keys used by stored data files and communications (email, network access, etc), protect in hardware Personally Identifiable Information, such as user IDs and passwords and protect passwords and credentials stored on drives\cite{urltcg}.

P2P technologies currently are used to construct distributed systems.  P2P has been widely used in file sharing, instant message, telephone communications, etc.  But, it is difficult to construct distributed systems like distributed operating systems.      There are two kinds of P2P technologies.   The un-structural P2P technology such as Freenet\cite{Clarke2000} using flooding way has shortage of heavy traffic and un-guaranteed search. The structural P2P technology using DHT such as Chord \cite{Stoica2001} and Pastry \cite{Rowstron2001} loses semantic meanings.   Semantic P2P networks  ( virtual hierarchical tree Grid organizations(VIRGO)) \cite{LHuangVirgo}\cite{LHuangP2P}\cite{agentH} can avoid the above disadvantages .  We use semantic P2P networks to construct distributed system ( DScloud )\cite{DScloudplatform} which can  do distributed search, parallel computing, etc.

JAVA is a programming language that is  intended to let application developers "write once, run anywhere"  and   can run on all platforms that support Java virtual machine.  With JAVA,  code mobility is also very easy to be implemented.  In distributed systems, in many cases , we need the processes of moving mobile codes across the nodes of a network.  JAVA P2P distributed systems involve many heterogenous environments and various people who are different interest. It is difficult to control users' behaves.  In order to  guarantee the correct  results of computing and to avoid  attacking by all kinds of attackers,   in JAVA P2P systems, how to check the integrity of the programs becomes a vital important issue.

We here present an  integration checker of JAVA  P2P distributed  System with auto source code composition .  It is more safe than traditional security methods.

The structure of this paper is as follows: section 2  describes the purpose of integrity checker of distributed system;  section 3 describes
framework for  integrity checker of JAVA distributed  system;   section 4 presents  implementation  of  integrity checker in DScloud platform;  section 5   analyzes the integrity check against attackers  and finally we give conclusions.

\section{The Purpose of Integrity Checker of Distributed  System }

For P2P distributed system, integrity checker of program is pre-requisite  service.    The purposes of integrity checker of distributed  system  are as the following:

\textbf{1.  against attacker to forge identification}

    In distributed systems forge identification can make the users believe the attackers are the forged users, which are harmful in instant message   \cite{p2pims}or information retrieve.  For example,  some person is framed by user A  who forges as user B with broadcasting false messages. Obviously, the person is harmed , and all people think   user B broadcast the message.  Obviously , wether the cheat is successful or not ,  all happens will be beneficial to user A , harmful to user B.  If the integrity of program can not  been guaranteed , the program can be changed to implement the forge identification very easily.  As P2P distributed systems with so many users, this case may happen very often.    In another case, the forge attacker can retrieve access control information of the other computers.  If there is no integrity checker, people are afraid  to use the program even in the case that the company  vows the system is safe.

\textbf{2.  against attacker to access the privacy data}

    In P2P distributed systems the node computers have many privacy data. In the cases such as file sharing and local search engines\cite{patent3}, we must limit the file directories to share and retrieve.   If the integrity of program can not  been guaranteed , the program can be changed to enlarge the scope of the directories even covering all the file system,  especially , when the program is developed by JAVA  which can be easily decompiled and modified.

\textbf{3.  against attacker to mislead the wrong results}

    In P2P distributed systems, the nodes are used to do parallel computing\cite{p2phighperformance}. To guarantee of  the integrity of program is most important. If the integrity of program can not  been guaranteed , the program can be changed to mislead the wrong results of the computing.  For example,   We have implemented  a distributed knapsack problem solution in our  DScloud platform, in which sub knapsack problems will be sent to P2P nodes to compute. If  the program in one of the nodes has changed, that will lead to get the wrong results.

\section{Framework for  Integrity Checker of JAVA Distributed  System}

    In a distributed systems developed by JAVA, integrity checker at least includes  a minimum of three components similar described in \cite{MikeBurmester}:

\textbf{1. Protected capabilities}   program and its required Jars , and JAVA virtual machine environment .

\textbf{2. Integrity measurement}    the process of obtaining metrics of
platform characteristics that affect the integrity (trustworthiness)
of a platform.

\textbf{3. Integrity reporting}   (i) to expose  properties  of integrity measurements, and (ii)
to attest to the integrity of  the measured  properties.

In P2P distributed systems which have NAT\cite{NAT} servers to provide the out IP and port, the servers maintains the integrity properties of the programs. Fig 1. shows  framework for  integrity checker of distributed  system  we presented.

\begin{figure}[h]
 \begin{center}
 \setlength{\unitlength}{1cm}
 \begin{picture}(10,6)
 \includegraphics[scale=.45]{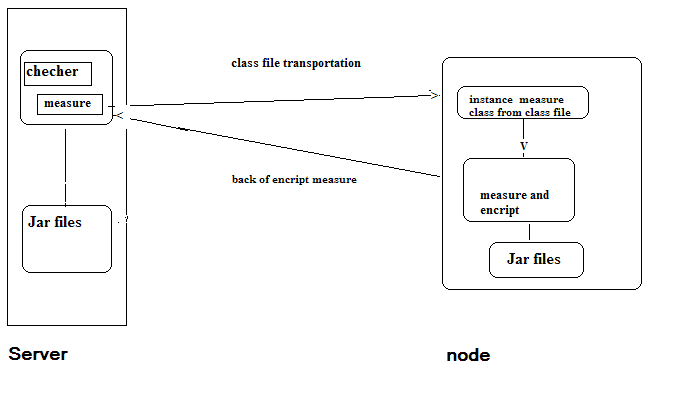}
 \end{picture}
 \caption{ Framework for  Integrity Checker of JAVA Distributed  System } \label{1}
 \end{center}
\end{figure}

In the framework of the integrity checker, the measurement and encrypt methods are produced by generating just-in-time source codes and compiling  by the server, and the program gets class file  from server and dynamically invokes instance of object of the class. So, every time every user, the measurement tool and encrypt methods are quite different,  this makes hacker impossible to bypass the integrity check process.

\section{ Implementation  of  Integrity Checker in DSclound Plateform}

The DScloud platform is developed by Hangzhou Domain Zones technology Co., LTD and Hangzhou Domain Search Networking Technology Co.,  LTD.

The DScloud platform is based on semantic P2P networks, which the nodes are virtually grouped as hierarchical classifications according to the semantic meanings as  the fig.2 shows.

\begin{figure}[h]
 \begin{center}
 \setlength{\unitlength}{1cm}
 \begin{picture}(10,8)
 \includegraphics[scale=.35]{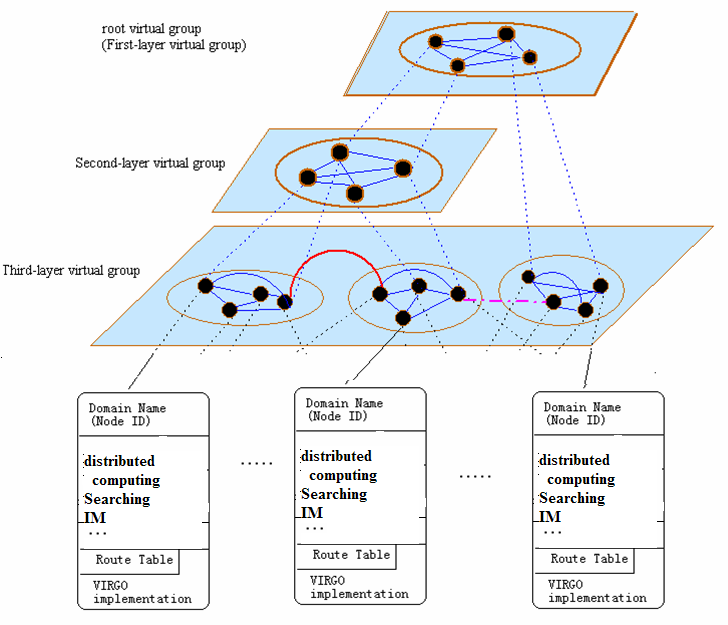}
 \end{picture}
 \caption{ Architecture  of DScloud Platform based on Semantic P2P Networks  } \label{1}
 \end{center}
\end{figure}

The DScloud platform includes one server program and many nodes' programs.  The server program functions NAT server, and integrity check, and storage of mobile codes. The components of the DScloud platform are as fig.3 shows.

\begin{figure}[h]
 \begin{center}
 \setlength{\unitlength}{1cm}
 \begin{picture}(10,8)
 \includegraphics[scale=.45]{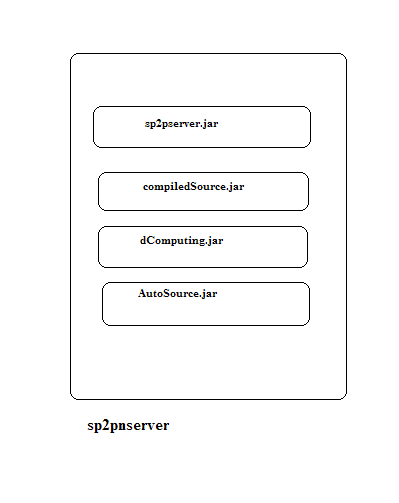}
 \end{picture}
 \caption{  The components of the DScloud platform Server } \label{1}
 \end{center}
\end{figure}

The AutoSource.jar includes Checker which checks the node's program. The checker includes auto source code composition which generates measurement tool and encrypt methods.

The integrity check protocol is as following :

\begin{verbatim}
Step 1  node's program sp2pen.jar sends the autoCheck message
        which indicates the node ID  to sp2pserver

Step 2  sp2pserver receives the autoCheck message , and invoke
        auto source  composition and generate autoCheck class
        file .

Step 3  node's program downloads autoCheck class file , and
        instances autoCheck class and invoke measurement tool
        to get integrity properties of the program, and uses
        the encrypt methods of autoCheck to  encrypt the
        properties such as file MD5 digests which are to be
        sent back to the server.

Step 4  The server uses the same  measurement tool and encrypt
        methods to get the integrity properties from the store
        files such as sp2pen.jar, and compares  these values
        with the received ones.

Step 5  if the values are the same, then records that the node
        is pass, otherwise, it is fail and the node is
        prohibited  to  get the next functions. In the mealtime,
        send the check status to the node.

Step 6  if the status the node received  is pass, it will
        continue to run, otherwise, it stops.



\end{verbatim}

\section{The Analysis of Integrity Check against Attackers  }

In sp2pen.jar, the code point for the process for integrity check  sends  autoCheck message and invokes the downloaded autoCheck class and get the returned status of integrity. Due to the JAVA's easy decompiling, the attackers can bypass  these steps and go to the next steps.  However, these steps are necessary to the server because  the server records  this node's integrity status;  that is, if these steps are removed, then server will not let the node program work further because there are vital necessary classes which are downloaded form server.

The autoCheck class which server generated  includes measure tools and encrypt methods and inner socket communications.  Because this  autoCheck class  is generated for the request of node, each node's request will generate different autoCheck class. The attackers can not cheat the server   because the attackers can not know the just-in-time measure tools and encrypt methods.

There are many  measure tools, but the MD5 digest is easy.  We can calculate the value of  sp2pen.jar with the slight modification of MD5 digest algorithm , and encrypt and sends the values to the server, the server uses the same  measure tools and encrypt methods to get the values against the sp2pen.jar file stored in the server.

\section{Conclusion}
 This paper presents a Integrity Checker of JAVA  P2P  distributed  system  with just-in-time auto source code composition.  Integrity of program itself and the system components of JAVA virtual machine are vital important for JAVA P2P distributed  systems against attackers, hackers, spies, cheaters conspirators, etc.   We here present a novel method using  just-in-time auto source code composition to generate just-in-time integrity measure tools and encrypts  of integrity reporting. We also analyze  integrity guarantees  against various kinds of attackers.
 By  the   effort of Hangzhou Domain Zones Technology Co., Ltd. and Domain Search Networking Technology Co., Ltd.  ,  we have implemented and use it in DSCloud platform.

\section*{Acknowledgements}
The paper is supported by the project "e-business Plateform via Virtual Community constructed from Semantic P2P Networks" supported by Zhejiang future science and Technology City, and ¡°Hangzhou Qinglan
Plan--scientific and technological creation and development
(No.20131831K99¡± supported by  Hangzhou scientific and technological
committee. The software copyrights is owned by Hangzhou
Domain Zones Technology Co., Ltd. and Domain Search Networking Technology Co., Ltd., and Chinese patent
applied is owned by Hangzhou Domain Zones Technology
Company and Domain Search Networking Technology Co., Ltd..

\end{document}